\documentclass[aip,jcp
 amsmath,amssymb,
 reprint,%
]{revtex4-2}

\usepackage{graphicx}
\usepackage{dcolumn}
\usepackage{bm}
\usepackage{comment}

\usepackage[utf8]{inputenc}
\usepackage[T1]{fontenc}
\usepackage{mathptmx}
\usepackage{etoolbox}
\usepackage{booktabs}
\usepackage{makecell}
\usepackage{dcolumn}
\usepackage{xcolor}

\makeatletter
\def\@email#1#2{%
 \endgroup
 \patchcmd{\titleblock@produce}
  {\frontmatter@RRAPformat}
  {\frontmatter@RRAPformat{\produce@RRAP{*#1\href{mailto:#2}{#2}}}\frontmatter@RRAPformat}
  {}{}
}%
\makeatother
\begin{document}


\title[Ehrenfest dynamics with PAW and LCAO]{Ehrenfest dynamics with localized atomic-orbital basis sets within the projector augmented-wave method}
\author{Vladimír Zobač}
\email[Corresponding authors: ]{vladimir.zobac@univie.ac.at, toma.susi@univie.ac.at}
\affiliation{%
University of Vienna, Faculty of Physics, Boltzmanngasse 5, 1090 Vienna, Austria
}%
\author{Mikael Kuisma}
\author{Ask Hjorth Larsen}
\affiliation{CAMD, Department of Physics, Technical University of Denmark, Lyngby 2800 Kgs., Denmark}
\author{Tuomas Rossi}
\affiliation{Department of Applied Physics, Aalto University, P.O. Box 11100, FI-00076 Aalto, Finland}
\affiliation{CSC – IT Center for Science Ltd., P.O. Box 405, FI-02101, Espoo, Finland}
\author{Toma Susi}
\affiliation{%
University of Vienna, Faculty of Physics, Boltzmanngasse 5, 1090 Vienna, Austria
}%

\date{\today}

\begin{abstract}
Density functional theory with linear combination of atomic orbitals (LCAO) basis sets is useful for studying large atomic systems, especially when it comes to computationally highly demanding time-dependent dynamics. We have implemented the Ehrenfest molecular dynamics (ED) method with the approximate approach of Tomfohr and Sankey within the projector augmented-wave code GPAW. We apply this method to small molecules as well as larger periodic systems, and elucidate its limits, advantages, and disadvantages in comparison to the existing implementation of Ehrenfest dynamics with a real-space grid representation. For modest atomic velocities, LCAO-ED shows satisfactory accuracy at a much reduced computational cost. This method will be particularly useful for modeling ion irradiation processes that require large amounts of vacuum in the simulation cell.
\end{abstract}

\maketitle

\section{\label{sec:intro}Introduction}

Molecular dynamics (MD) in conjunction with density functional theory (DFT)~\cite{marx_hutter_2009} is a robust approach for simulating relatively large systems from first principles~\cite{Smith2020}. It provides efficient means for theoretical investigations into atomic and electronic properties, supporting experimental observations~\cite{Miyamoto2018}. Further describing the time-dependent quantum-mechanical evolution of the electron density along classical nuclear dynamics enables a proper understanding of nanoscale phenomena typically accompanied by electronic excitation such as charge transfer, photoabsorption, and irradiation of solids, liquids, and two-dimensional (2D) nanomaterials. Numerous semi-classical methods have been developed and applied in this context~\cite{Tully_1998, Yarkony2011, Worth_2008, Gaigeot2010, Tully_2012, Tavernelli2015, Ma2016, Churchod2018,  Yost2019, Zhang2020}. The Ehrenfest method is a natural extension of ground-state DFT/MD to include time-dependent electron dynamics, which has been successfully applied to describe the electronic stopping of projectiles in bulk materials~\cite{Ullah2015, Halliday2019}.

Recent extensive experimental studies have explored collisions involving non-trivial charged ion projectiles and 2D nanomaterials~\cite{Wilhelm2015, Zhang2012, Wilhelm2022, Li2022, Kononov2022, Niggas2023}, with promising applications in areas including helium ion microscopy or controlled nanopore fabrication. The complex interaction between ions and the target materials in these experiments lead to processes such as pore creation, electron emission, atom sputtering, projectile capture, and more. Simulations of such experiments often require large supercells with vacuum and extended simulation times, placing considerable demands on computational methods and limiting the types of systems that can be studied.

The technical implementation of the Ehrenfest method varies depending on the chosen flavor of DFT. Typically, DFT provides a potential energy surface (PES) from which forces acting on nuclei can be computed. The shape of the effective PES is given by electrons, whose degrees of freedom are propagated by the time-dependent Schrödinger equation, and nuclei that are evolved by classical Newtonian mechanics.

Kohn--Sham DFT can employ various representations of the electronic wave functions, each suited to different applications. Common representations include basis sets such as plane-waves, real-space grids, linear combinations of atomic orbitals (LCAO), and Gaussian functions.~\cite{Castro2006, garcia_siesta_2020, Kuhne2020, Mortensen2023} Further, for practical reasons, most DFT approaches require a computationally efficient representation of inner electrons and nuclei. Key methodologies in this context include the utilization of pseudopotentials and the projector augmented-wave method (PAW)~\cite{Blochl1994}.

The basis set not only impacts the accuracy and efficiency of the calculation, but also determines how quantities such as forces and densities are computed as well as the algorithms for propagating the electronic degrees of freedom. Specifically, basis sets that are parametrically independent of the nuclear positions such as plane waves or real-space grid representations, differ in their approach from those that do depend on the nuclear position, such as localized atomic orbitals or Gaussian functions. These distinct flavors of basis sets require different strategies for electron propagation and force calculations, especially in the context of molecular dynamics~\cite{Kunert2003, Castro2006}.

The latter localized approach, while computationally more efficient, adds velocity-dependent terms into the Hamiltonian and forces, which have been identified as mathematically equivalent to a connection in differential geometry~\cite{Artacho2017}. With this approach, the integration of electronic degrees of freedom can be interpreted as a gauge potential in the Hamiltonian, appropriately named gauge-potential (GP) integration~\cite{Artacho2017}. Another more approximate method for integrating electronic degrees of freedom was proposed by Tomfohr and Sankey (TS)~\cite{Tomfohr2001}, which introduces a Löwdin orthonormalization step into the integration procedure. 

In the Siesta package~\cite{soler_siesta_2002,garcia_siesta_2020}, which utilizes pseudopotentials and LCAO, the GP integrator was found to be accurate even for higher velocities, whereas the TS integrator showed reduced accuracy~\cite{Halliday2021}. Unfortunately, the GP integrator is very difficult to implement within the PAW method as the projector functions also depend on the nuclear positions. However, the TS approximation is relatively easier to adapt, which is what we will describe in this contribution.

The Ehrenfest method was implemented~\cite{Ojanpera_2012} in the real-space finite-difference wavefunction representation (FD-PAW-ED) in of the projector-augmented-wave code GPAW~\cite{enkovaara_electronic_2010,Mortensen2023}, and successfully applied to study the electronic stopping of ions~\cite{ojanpera_electronic_2014} amongst other phenomena~\cite{Mortensen2023}. In this study, we compare results obtained from that method to those from our recently implemented PAW Ehrenfest method utilizing LCAO (LCAO-PAW-ED) following the strategy proposed by TS~\cite{Tomfohr2001}, which has not been previously studied. To propagate the electronic degrees of freedom, we employ the semi-implicit Crank-Nicholson scheme~\cite{Mortensen2023}, while the nuclei are treated as classical particles. Our comparative analysis applies these methods to diverse systems including NaCl, CH$_2$NH$_2^+$, and a periodic cell of monolayer graphene impacted by H and H$^+$, allowing us to elucidate the limitations of the methods and to discuss their respective advantages as well as disadvantages.

\subsection{\label{sec:classical}Classical equations of motion for nuclei}

The temporal evolution of the classical degrees of freedom of the nuclei, denoted by $a$, is governed by Newton's equation of motion:
\begin{equation}
M_a \ddot{\mathbf{R}}_a = \mathbf{F}_a,
\end{equation}
where $M_a$ represents the mass of nucleus $a$, $\mathbf{R}_a$ its position vector, and $\mathbf{F}_a$ the forces acting on it. To advance the nuclear dynamics, we employ the well-established classical velocity-Verlet propagator as previously implemented~\cite{Ojanpera_2012}. This propagator calculates the first step for nuclei under the assumption of a constant Hamiltonian. Subsequently, it computes the averaged Hamiltonian using both the predicted and the previous step, utilizing it for the subsequent propagation of the electrons.

The forces acting on the nuclei are determined as the negative derivatives of the system's electronic energy, i.e., $\mathbf{F}_a = -{d E_{\mathrm{el}}}/{d \mathbf{R}_a}$. To maintain consistency with prior implementations, we adopt the original approach for calculating these forces~\cite{Larsen2009}. This choice is motivated by the desire to circumvent issues arising from the dynamic repositioning of basis functions and projectors, which would necessitate the inclusion of additional terms in the forces to ensure proper energy conservation~\cite{Kunert2003, Qian2006, Ojanpera_2012}. Although velocity-dependent terms are neglected, which may only be valid within specific velocity ranges, achieving a comprehensive description of forces would demand a more intricate effort beyond the scope of the present work. We also note that the PAW approach adds additional cross terms that make it more difficult to implement than prior pseudopotential approaches~\cite{garcia_siesta_2020}.

\subsection{\label{sec:quantum}Quantum equations of motion for electrons}
We then derive the relevant quantum equations of motion following Ojanperä~\cite{ojanpera_simulating_2016}. The general formulation of the time-dependent Kohn-Sham (KS) equation for a system of atoms assuming static nuclei is
\begin{eqnarray}
\label{eq:KS-gen}
i \frac{\partial}{\partial t} \psi_k(\mathbf{r}, t)=\hat{H} \psi_k(\mathbf{r}, t),  
\end{eqnarray}
where $\psi_k(\mathbf{r}, t)$ are single-particle KS electronic states and $n(\mathbf{r}, t)=\sum_k\left|\psi_k(\mathbf{r}, t)\right|^2$ is the electronic density. The PAW formalism introduces pseudized all-electron Kohn-Sham wavefunctions $\psi_n(\mathbf{r}, t)=\hat{\mathcal{T}} \tilde{\psi}_n(\mathbf{r}, t)$, and considering that nuclei are no longer static, the equation above transforms into
\begin{eqnarray}
\label{eq:KS-PAW}
    i \tilde{S} \frac{\partial \tilde{\psi}_n}{\partial t}=[\tilde{H}(t)+\tilde{P}(t)] \tilde{\psi}_n(\mathbf{r}, t)
\end{eqnarray}
where the operator $\tilde{P}(t)$ derived for the PAW approach in Ref.~\onlinecite{Qian2006} as
\begin{eqnarray}
\tilde{P}(t)=-i \hat{\mathcal{T}}^{\dagger} \frac{\partial \hat{\mathcal{T}}}{\partial t}
\end{eqnarray}
captures the motion of the projector functions. The operator $\hat{\mathcal{T}}$ is the PAW transformation operator~\cite{Mortensen2023}. Here, the PAW overlap operator is defined as $\tilde{S}=\hat{\mathcal{T}}^{\dagger} \hat{\mathcal{T}}$ and the PAW Hamiltonian including any external time-dependent potential as $\tilde{H}(t)=\hat{\mathcal{T}}^{\dagger} \hat{H}(t) \hat{\mathcal{T}}$.

In the LCAO approach, the wavefunction is expanded in a set of localized orbitals $\Phi_\mu$ as
\begin{eqnarray}
\label{eq:LCAO-exp}
\left|\tilde{\psi}_n\right\rangle=\sum_\mu c_{n \mu}\left|\Phi_\mu\right\rangle
\end{eqnarray}
with expansion coefficients $c_{n \mu}$ where $n$ indexes the KS states and $\mu$ the localized orbitals. Substituting expansion (\ref{eq:LCAO-exp}) into the PAW-TDKS equation (Eq.~\ref{eq:KS-PAW}) yields
\begin{eqnarray}
  \label{eq:KS-gen-PAW-LCAO}
i \tilde{S} \sum_\mu\left(\frac{\partial c_{n \mu}}{\partial t}|\Phi_\mu\rangle+c_{n \mu} \frac{\partial|\Phi_\mu\rangle}{\partial t}\right)=(\tilde{H}+\tilde{P}) \sum_\mu c_{n \mu}|\Phi_\mu\rangle.
\end{eqnarray}
By further operating from the left with $\left\langle\tilde{\psi}_n\right|=\sum_\nu c_{n 
\nu}^*\left\langle\Phi_\nu\right|$ and rearranging the terms, one obtains the matrix equation
%
%
\begin{eqnarray}
  \label{eq:KS-PAW-LCAO-final}
i \sum_\mu \tilde{S}_{\nu \mu} \frac{\partial c_{\mu n }}{\partial t} = \sum_\mu(\tilde{H}_{\nu \mu}+\tilde{P}_{\nu \mu}) c_{\mu n} -i \sum_\mu {G_{\nu \mu}}\cdot {\mathbf{v}}^{\nu} {c_{\mu n }}
\end{eqnarray}
where $\mathbf{c}_n$ are the expansion coefficients of electronic state $n$.

The matrices $S_{\nu \mu}=\left\langle\Phi_\nu|\tilde{S}| \Phi_\mu\right\rangle$ describe the overlap and $ H_{\nu\mu}=\left\langle\Phi_\nu|\tilde{H}|\Phi_\mu\right\rangle$ the Hamiltonian in the LCAO basis. The matrix $G_{\nu \mu} = \left\langle\Phi_\nu | \hat S | \boldsymbol\nabla\Phi_\mu\right\rangle$ is proportional to the velocity of the atom $\mathbf v^\nu$ hosting basis function $\nu$. The $\mathbf{c}_n$ elements of the matrices in Eq.~\ref{eq:KS-PAW-LCAO-final} are calculated using the semi-implicit Crank-Nicholson scheme~\cite{Kuisma2015}. Notably, the matrices $\mathbf{P}$ and $\mathbf{G}$ express the connection in the manifold for the projectors and the LCAO basis, respectively.

To approximate the transformation of the basis in a square bracket we perform Löwdin orthogonalization proposed by Tomfohr and Sankey~\cite{Tomfohr2001}:
\begin{eqnarray}\label{eq:TS}
\mathbf{c}_n\left(t_0+\Delta t\right)=\mathbf{S}^{-\frac{1}{2}}\left(t_0+\Delta t\right) \mathbf{S}^{\frac{1}{2}}\left(t_0\right) \tilde{\mathbf{c}}_n\left(t_0+\Delta t\right),
\end{eqnarray}
which transforms the time-dependent $\mathbf{c}_n$ by the square root of the overlap matrix whenever the atoms are moved from their position. When atoms move, the Hamiltonian and coefficients change stepwise. This discontinuity at the boundary between time steps is addressed by Eq.~\ref{eq:TS} and can be considered as fulfilling the condition of wavefunction continuity.

\section{Simulating molecules and periodic systems}
We will compare the implemented LCAO-PAW-ED method to the earlier FD-PAW-ED method on two small molecules, NaCl and CH$_2$NH$_2^+$, as well as atomic H and a H$^+$ ion colliding with a periodic graphene monolayer. We use the $dzp$ basis set for graphene and molecules and $sz$ basis set for the H projectile, and a grid sampling of 0.2~\AA\ with the PBE exchange-correlation functional. The C atoms were represented using the standard datasets of GPAW where 1$s$ electrons are included only as frozen core states. Although it is possible to run simulations with PAW datasets that include 1$s$ as a valence state, we do not expect this to be necessary for the relatively deep C 1$s$ level and the impact parameters we consider here~\cite{ojanpera_simulating_2016}. 

For molecules, we use a rectangular simulation box, and for graphene choose a 6$\times$6 periodic supercell with $\Gamma$-point sampling of \textbf{k}-space, as simulations with 5 \textbf{k}-points ($~\Gamma~\Sigma~M~K~\Lambda$) showed insignificant differences. To prepare the ionic state of H$^+$, we used charge-constrained density functional theory~\cite{melander_implementation_2016}, where an extra potential is applied in the site of the projectile in order to remove charge from this area. The LCAO method is advantageous in this context due to its lower computational cost, particularly for systems with significant vacuum. However, when computationally feasible, FD methods might offer more consistent accuracy for high charge states.

\subsection{Vibrations of NaCl}

We choose the diatomic NaCl molecule as the first test case to demonstrate the correspondence between LCAO-PAW-ED and FD-PAW-ED, and to compare them to Born-Oppenheimer (BO) MD simulations. A kinetic energy of 1~eV was applied antiparallel to the direction of the Na--Cl bond to induce molecular stretching vibrations in the adiabatic regime. The maximum resulting velocity of atoms was $\sim$0.02~\AA/fs, with an equilibrium distance of 2.35~\AA. The total simulation time was 300~fs, and timesteps between 10 and 1000~as were tested. 

\begin{figure}
    \centering
    \includegraphics[width=0.48\textwidth]{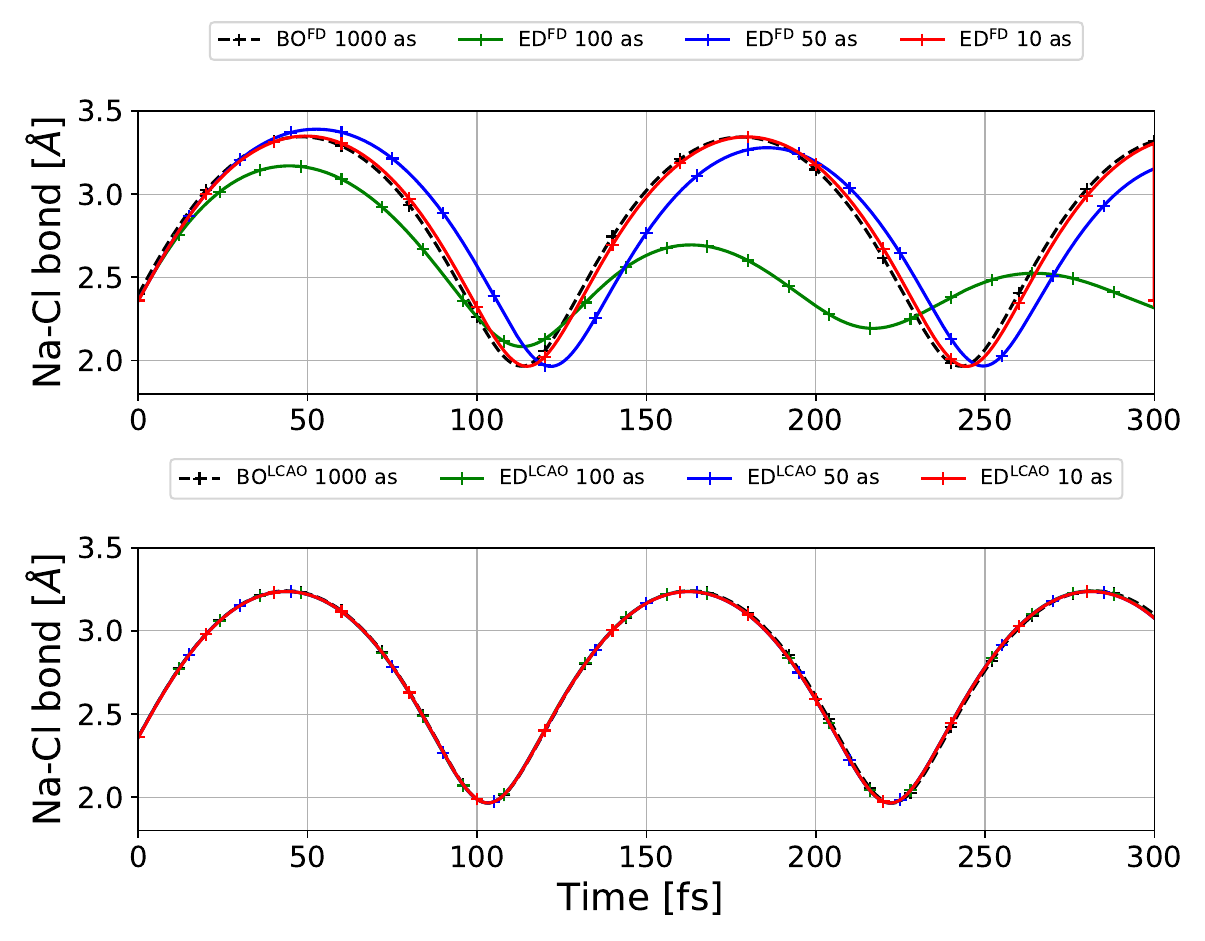}
    \caption{Na--Cl bond-length oscillation comparison for BO-MD, FD-PAW-ED, and LCAO-PAW-ED. A series of three different time steps was used with an excitation energy set to 1 eV. The upper panel shows how the FD-PAW-ED method requires a smaller timestep down to 10 as for convergence with respect to BO, whereas LCAO-PAW-ED remains accurate for much longer steps (lower panel).}
    \label{fig:Na_Cl_bond}
\end{figure}

The bond oscillation in each method should be similar to BO dynamics. The oscillation period was 132.2~fs for FD-PAW-ED (BO-MD 130.1~fs) and 119.6~fs for LCAO-PAW-ED (BO-MD 119.03~fs), with a small difference due to the chosen basis set. In the case of FD-PAW-ED, the accuracy of the calculation is more sensitive with respect to the timestep, while for LCAO-PAW-ED, a very similar oscillation period is observed for all chosen steps as shown in Figure~\ref{fig:Na_Cl_bond}. 

In terms of energy conservation, the change in total energy for FD-PAW-ED increases from 0.37 to 3.45 eV when the timestep is increased from 10 to 100 as. By contrast, for LCAO-PAW-ED the total energy is conserved within 4.5 meV for the same range of timesteps. The greater sensitivity of FD-PAW-ED to the timestep may be attributed to the grid representation of wavefunctions where $N^3$ (where N is the number of grid points in each direction) points are evolved in time, producing a much greater numerical error. A longer stable timestep points towards the possibility of even greater savings of computing time when using LCAO-PAW-ED in certain simulation settings.

\subsection{Isomerisation of CH$_2$NH$_2^{+}$}

Protonated formaldimine is known to be a molecule that is subject to photoisomerisation. Various states can be populated by photoexcitation, and subsequently, corresponding decay pathways are followed. Numerous theoretical molecular dynamics studies employing various methods have been conducted to describe the relaxation process of the molecule following photoexcitation~\cite{Barbatti_2006, Tavernelli2010, Ojanpera_2012}. 

\begin{figure}[t]
    \centering
    \includegraphics[width=0.47\textwidth]{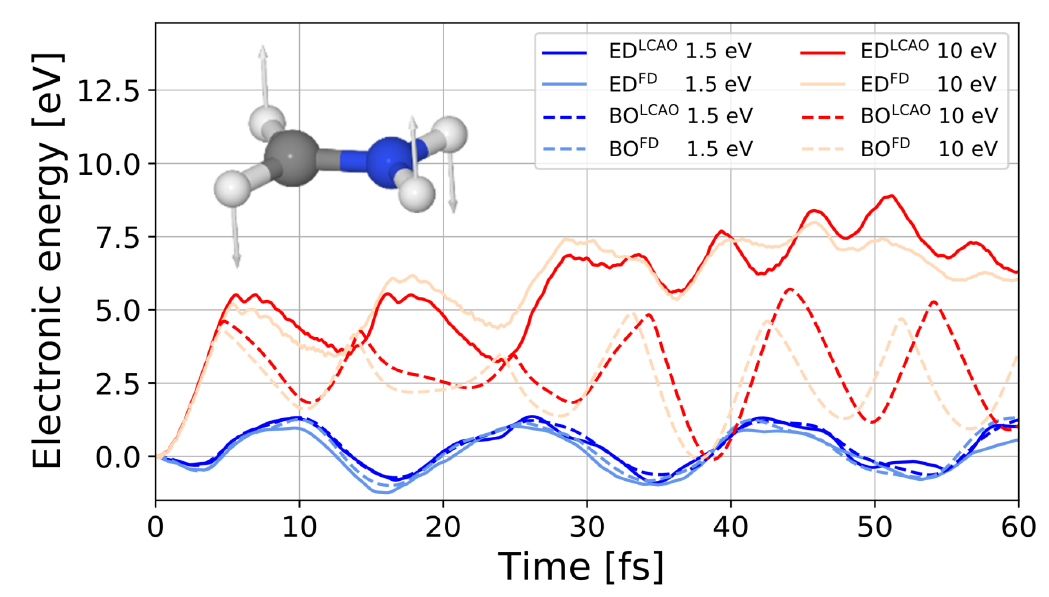}
    \caption{Potential-energy surface along the simulation trajectory for the CH$_2$NH$_2^{+}$ molecule with excitation energies of 1.5 and 10~eV. The initial velocity was set in the direction of the nonadiabatic coupling vector depicted in the figure.}
    \label{fig:PES_molecule}
\end{figure}

Instead of conventional photoexcitation, we excited the molecule by imparting velocities in the direction of the nonadiabatic coupling vector (as shown in the inset of Figure~\ref{fig:PES_molecule}) between the first excited $S_{1}(\pi \rightarrow \pi^*)$ state and the second excited $S_{2}(\pi \rightarrow \sigma^*)$ state. This approach activates the twist mode, propelling the molecule toward the conical intersection. Two activation energy levels were employed: $E_1 = 1.5$~eV to simulate the molecule's dynamics far from the conical intersection in the adiabatic regime, and $E_2 = 10$~eV to emulate the nonadiabatic effects of the photoisomerization process. The integration time step was 10 as.

\begin{figure}[b]
    \centering
    \includegraphics[width=0.47\textwidth]{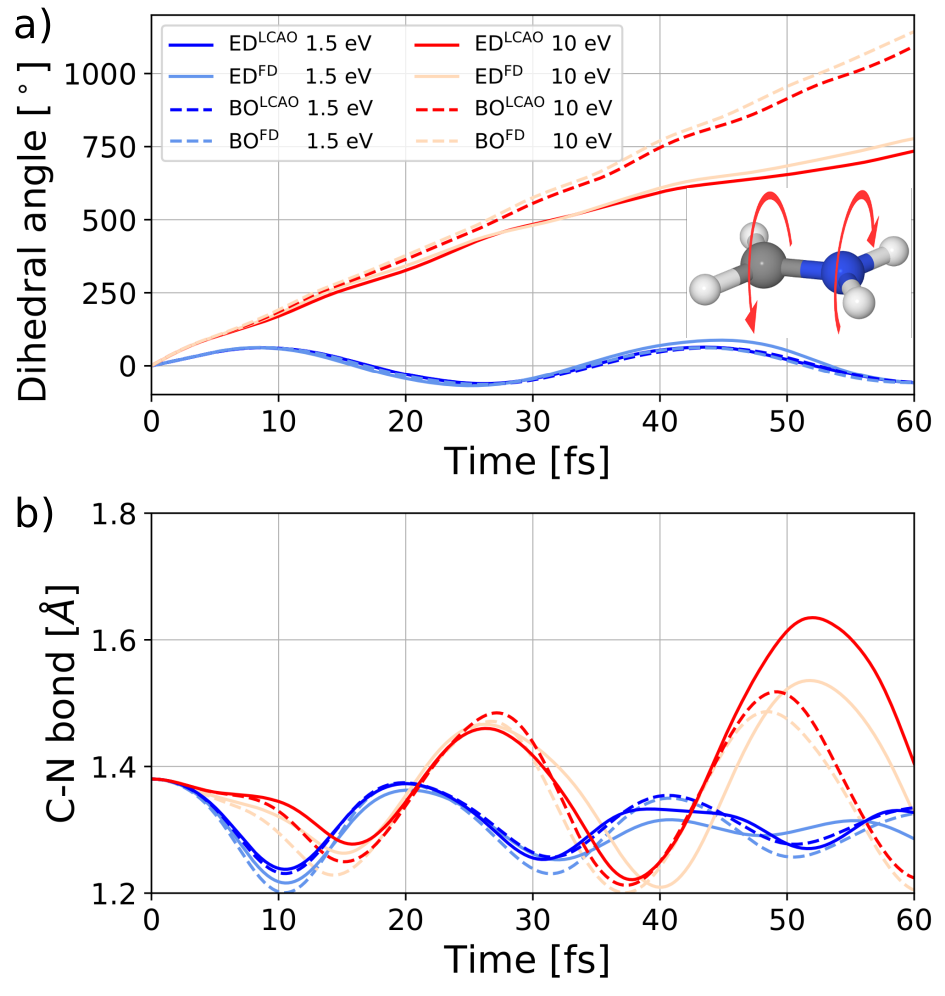}
    \caption{a) Dihedral angle between the HCNH atoms. During the 10~eV
    simulation, angles beyond 180° should be understood as cumulative, i.e., atoms rotate multiple times around the bond. b) The C--N bond length for the excited CH$_2$NH$_2^{+}$ molecule.}
    \label{fig:CH2NH2_dih_bond}
\end{figure}

The simulation results are depicted in Figures~\ref{fig:PES_molecule} and \ref{fig:CH2NH2_dih_bond}. Figure~\ref{fig:PES_molecule} illustrates the potential energy profile of the molecule. The lower-energy excitation at 1.5~eV induces a periodic stable twist oscillation of $65^\circ$ and C--N bond oscillation of 0.2~\AA\ in magnitude (Fig.~\ref{fig:CH2NH2_dih_bond}). Both LCAO- and FD-PAW-ED methodologies exhibit similar oscillations as BO simulations, suggesting that the molecule is in an adiabatic region. A markedly distinct behavior emerges when the same mode is excited with an energy of 10~eV: the molecule is propelled towards a conical intersection, reaching it within the first 10~fs. During the simulation time, the NH$_2$ and CH$_2$ groups undergo a rotation of $750^\circ$ before stabilizing around the perpendicular geometry corresponding to the conical intersection. The twisting of the molecule is accompanied by an increasing oscillation of the C--N bond, which may lead to dissociation.

By contrast, in BO simulations the CH$_2$ and NH$_2$ groups rotate without stabilization after passing the conical intersection (red and orange dashed lines in Figure~\ref{fig:CH2NH2_dih_bond}), whereas both FD- and LCAO-PAW-ED yield very similar results. It is however important to acknowledge that the realistic photochemistry of the molecule is not fully captured, as Ehrenfest dynamics in general is a mean-field theory. Nevertheless, it will be highly useful for simulating electronic stopping and collisions of projectiles, as will be demonstrated next.

\subsection{Irradiation of graphene with H and H$^{+}$}

\begin{figure}
    \centering
    \includegraphics[width=0.5\textwidth]{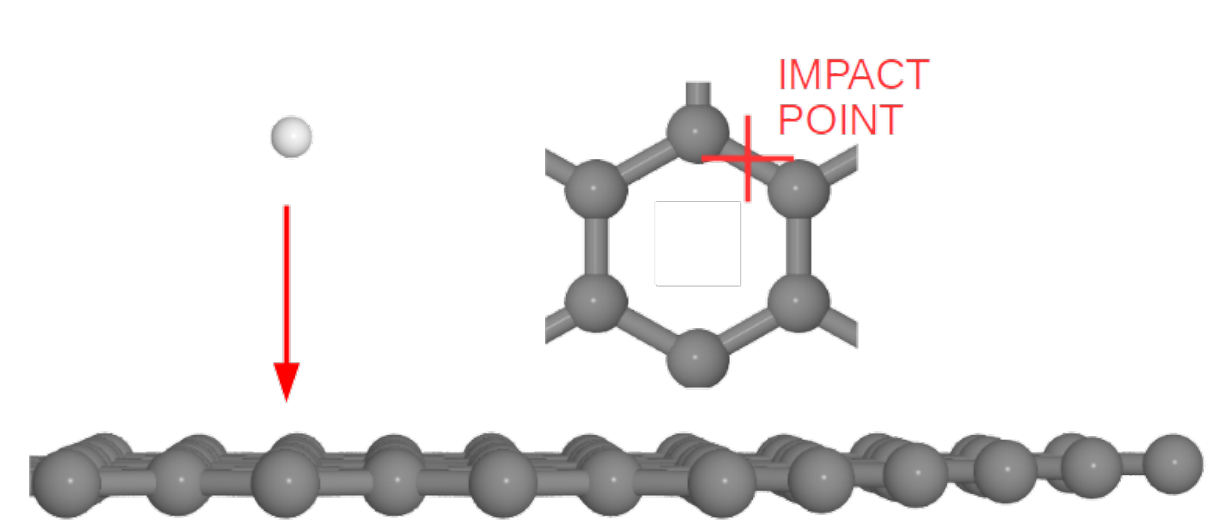}
    \caption{Graphene impact geometry with the H projectile.}
    \label{fig:H_graphene_scheme}
\end{figure}

To further demonstrate the capabilities and also drawbacks of the LCAO-PAW-ED approach, we simulate the irradiation of graphene with atomic hydrogen and hydrogen ion projectiles. This is a geometry that inherently requires a significant amount of vacuum, and which may therefore particularly benefit from a localized basis set. The impact parameter was selected in the middle of the C--C bond instead of the more typical hexagon-center to induce a greater interaction (Fig.~\ref{fig:H_graphene_scheme}). We compare neutral H with the H ion to elucidate the difference between a projectile that carries an electron (thus two parts of the system holding electrons) and a projectile that initially represents only a classical point charge.

\begin{figure}[b!]
    \centering
    \includegraphics[width=0.52\textwidth]{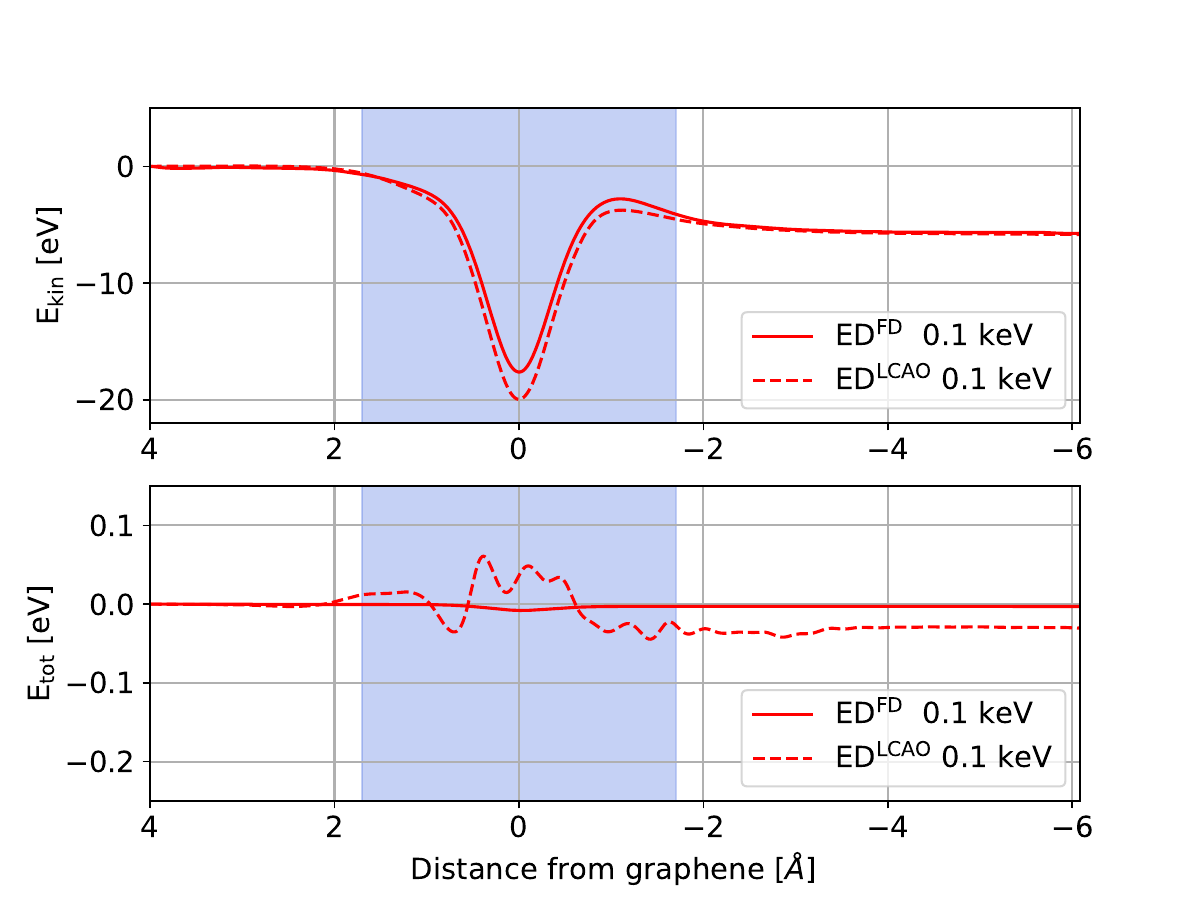}
    \caption{Kinetic energy loss and total energy conservation for 0.1~keV H, comparing LCAO- and FD-PAW-ED. The blue-shaded region represents van der Waals radius of carbon atoms in the target graphene monolayer.}
    \label{fig:Ekin_Hsz}
\end{figure}

\begin{figure}[b!]
    \centering
    \includegraphics[width=0.52\textwidth]{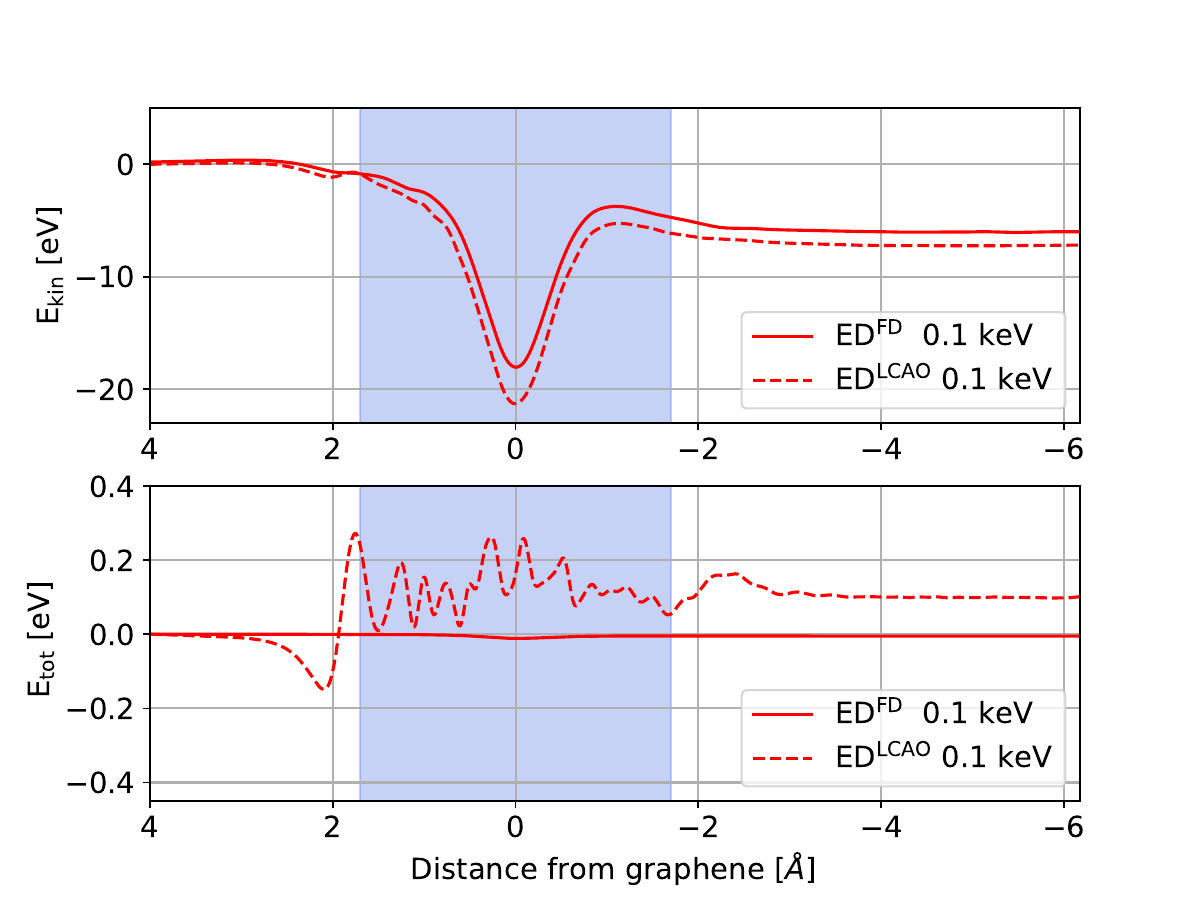}
    \caption{Kinetic energy loss and total energy conservation for 0.1~keV H$^+$, comparing LCAO- and FD-PAW-ED. The blue-shaded region represents van der Waals radius of carbon atoms in the target graphene monolayer.}
    \label{fig:Ekin_Hsz+}
\end{figure}

To test the sensitivity of LCAO-PAW-ED on the velocity of the projectile, we performed a set of calculations for  kinetic energies ranging from 0.1 to 10~keV. The timestep was conservatively 5~as for all simulations and the total simulation time was around 100~fs. The projectile was initially placed 4~\AA\ above the graphene plane. Representative plots of kinetic energy ($E_\mathrm{kin}$) loss and total energy ($E_\mathrm{tot}$)  conservation for a 0.1~keV kinetic energy (corresponding to a velocity of 1.26~\AA/fs) are depicted in Figures~\ref{fig:Ekin_Hsz} and \ref{fig:Ekin_Hsz+} for H and H$^+$ projectiles respectively. The differences in kinetic and total energy for other velocities are presented in Tables \ref{tab:table_Hsz} and \ref{tab:table_H+sz}.

The projectile experiences a slight deceleration before reaching the van der Waals distance from the atoms in graphene (indicated by the blue area in Figures~\ref{fig:Ekin_Hsz} and \ref{fig:Ekin_Hsz+}). It reaches its minimum kinetic energy at the graphene plane, with the LCAO-PAW-ED approach showing a slightly lower minimum compared to the FD-PAW-ED method. Although as a localized basis, LCAO may be affected by the basis set superposition error, we believe that the differences are mainly caused by the choice of Kohn-Sham states representation. After passing through the graphene layer, the projectile is accelerated again but does not regain the original kinetic energy it had prior to the collision.

\begin{figure}
    \centering
    \includegraphics[width=0.5\textwidth]{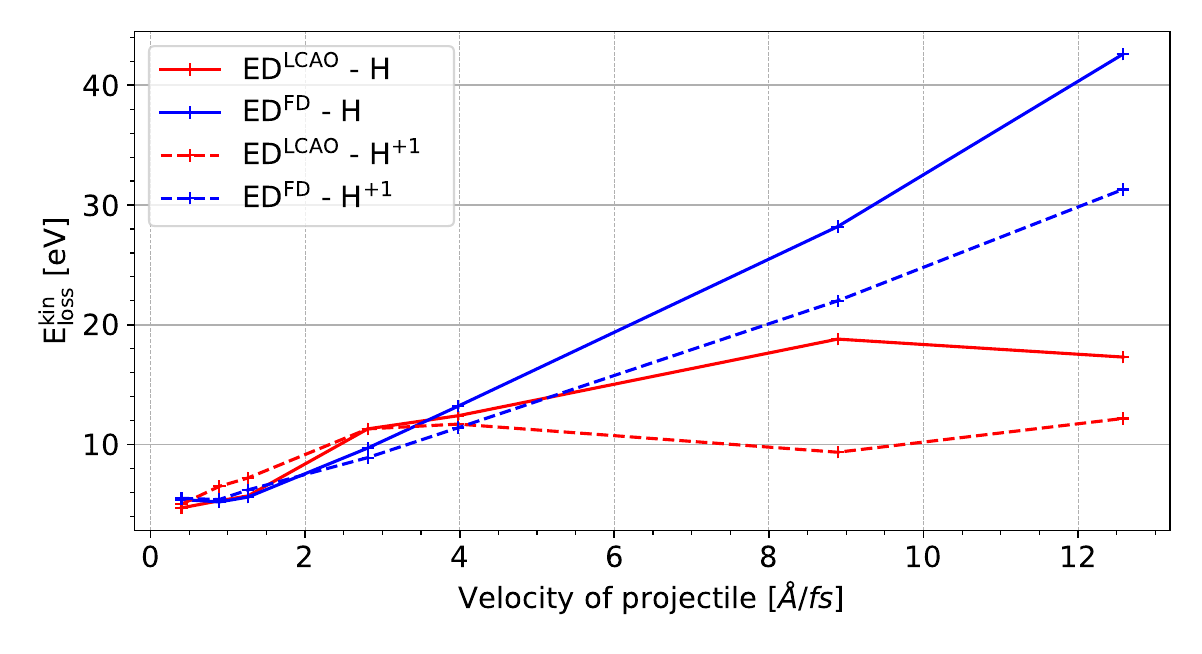}
    \caption{Kinetic energy loss of H projectile for different velocities.}
    \label{fig:Ekin_lossH_all}
\end{figure}

Kinetic energy loss of the projectile due to electronic stopping is very similar for FD- and LCAO-PAW-ED for 0.1~keV H$^+$ as can be seen from Tables~\ref{tab:table_Hsz} and \ref{tab:table_H+sz}, and similar to other methods~\cite{Kononov2021}. The difference between FD and LCAO can be attributed to the basis set. We see a slight acceleration of the projectile for FD-PAW-ED at the beginning of the simulation (by up to 0.2~eV), while for LCAO-PAW-ED the effect is missing, presumably due to the greater localization of the basis. The kinetic energy loss starts to notably deviate for 5~keV (8.89~\AA/fs) and above due to the velocities being too high to be accurately described by the TS-integrator in LCAO-PAW-ED (Fig.~\ref{fig:Ekin_lossH_all}). However, even for lower velocities, energy conservation depends on velocity and is also worse (0.011--0.372~eV) than for FD-PAW-ED (all within a few meV).

The projectile was reflected back from the C--C bond for a kinetic energy of 10~eV. In other cases, the projectile passed through the graphene layer without any structural changes. For both projectiles, the greatest violations of total energy conservation $E_{\mathrm{max}}^{H^{+}}$ and $E_{\mathrm{max}}^{H}$ appear when the projectile is passing the graphene layer. These unphysical oscillations of total energy are on order of a few tenths of an eV in the case of LCAO-PAW-ED, but only a tenth of a meV in the case of FD-PAW-ED (Tables~\ref{tab:table_Hsz} and \ref{tab:table_H+sz}). For higher velocities, the oscillations are increased also for FD-PAW-ED, which can influence the dynamics, and is likely due to charge sloshing (electronic density redistribution overcompensation from one self-consistency step to the next).

\begin{table}[t!]
    \centering
    \caption{Kinetic energy loss and total energy conservation for LCAO- and FD-PAW-ED for neutral H. $E^{\mathrm{kin}}$ is the initial kinetic energy of the projectile and $v$ the corresponding velocity, $E^{\mathrm{kin}}_{\mathrm{loss}}$ the kinetic energy lost to electronic stopping, $\Delta E$ the change in total energy (positive values correspond to final energy being lower than the initial), and $E_{\mathrm{max}}$ the maximum violation of energy conservation during the simulation. (Note the $\times10^{-2}$ factor for the last column.)}
    \label{tab:table_Hsz}
    \begin{ruledtabular}
    \begin{tabular}{*{5}{>{\centering\arraybackslash}p{1.6cm}}}
        \multicolumn{1}{c}{$E^{\mathrm{kin}}$ ($v$)} & \multicolumn{2}{c}{$E^{\mathrm{kin}}_{\mathrm{loss}}$ [eV]} & \multicolumn{2}{c}{$\Delta E\,(E_{\mathrm{max}})$ [eV]}\\
        \multicolumn{1}{c}{[keV (\AA/fs)]} & LCAO & FD & LCAO & FD ($\times10^{-2}$)\\
        \hline
        0.01 (0.40) & 4.7 & 5.4 & 23(38)$\times10^{-4}$  & 0.65 (0.52)\\
        0.05 (0.89) & 5.3 & 5.2 & 0.01(0.03)  & 0.31 (0.79)\\
        0.1 (1.26) & 5.7 & 5.6  & 0.03(0.06)  & 0.29 (0.81)\\
        0.5 (2.81) & 11.3 & 9.7 & 0.08(0.38) & 0.41 (0.87)\\
        1.0 (3.98) & 12.4 & 13.2 & -0.17(1.07) & 0.54 (0.92)\\
        5.0 (8.89) & 18.8 & 28.2 & -5.32(12.74) & 1.09 (1.43)\\
        10.0 (12.58) & 17.3 & 42.6 & -17.2(28.56) & 0.08 (1.04)\\
    \end{tabular}
    \end{ruledtabular}
\end{table}
\begin{table}[t!]
    \centering
    \caption{Kinetic energy loss and total energy conservation for LCAO- and FD-PAW-ED for H$^+$. $E^{\mathrm{kin}}$ is the initial kinetic energy of the projectile and $v$ the corresponding velocity, $E^{\mathrm{kin}}_{\mathrm{loss}}$ the kinetic energy lost to electronic stopping, $\Delta E$ the change in total energy (positive values correspond to final energy being lower than the initial), and $E_{\mathrm{max}})$ the maximum violation of energy conservation during the simulation. (Note the $\times10^{-2}$ factor for the last column.)}
    \label{tab:table_H+sz}
    \begin{ruledtabular}
    \begin{tabular}{*{5}{>{\centering\arraybackslash}p{1.6cm}}}
        \multicolumn{1}{c}{$E^{\mathrm{kin}}$ ($v$)} & \multicolumn{2}{c}{$E^{\mathrm{kin}}_{\mathrm{loss}}$ [eV]} & \multicolumn{2}{c}{$\Delta E\,(E_{\mathrm{max}})$ [eV]}\\
        \multicolumn{1}{c}{[keV (\AA/fs)]} & LCAO & FD & LCAO & FD ($\times10^{-2}$)\\
        \hline
        0.01 (0.40) & 5.0 & 5.5 & -0.01(0.05)  & 0.91 (0.90)\\
        0.05 (0.89) & 6.5 & 5.4 & -0.08(0.16)  & 0.51 (1.07)\\
        0.1 (1.26) & 7.2 & 6.2  & -0.10(0.27)  & 0.49 (1.13)\\
        0.5 (2.81) & 11.3 & 8.9 & -0.43(1.03) & 0.55 (1.26)\\
        1.0 (3.98) & 11.7 & 11.4 & -1.00(2.18) & 0.77 (1.39)\\
        5.0 (8.89) & 9.35 & 22.0 & -12.1(19.05) & 1.51 (2.13)\\
        10.0 (12.58) & 12.17 & 31.3 & -16.0(36.57) & 0.65 (2.43)\\
    \end{tabular}
    \end{ruledtabular}
\end{table}

\begin{figure}
    \centering
    \includegraphics[width=0.5\textwidth]{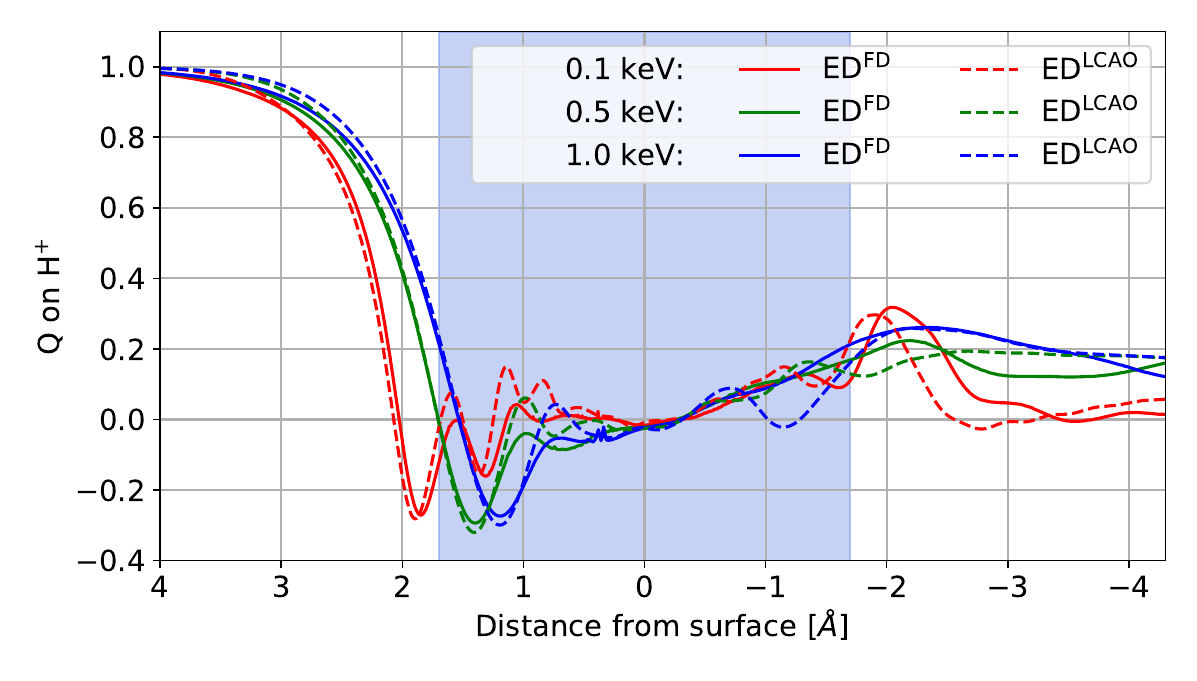}
    \caption{Hirshfeld charge on the H$^+$ projectile. The blue shaded region represents van der Waals radius of carbon.}
    \label{fig:Charge_H}
\end{figure}

Studying the neutralization of charged projectiles is one important use for these simulations. The Hirshfeld charge localized on the $H^{+}$ projectile is depicted in Figure~\ref{fig:Charge_H} and corresponds qualitatively to earlier published methods~\cite{Kononov2022}. For kinetic energies below 1~keV, the projectile is neutralized before it reaches the vdW radius of carbon atoms in graphene (Fig.~\ref{fig:Charge_H}). Both LCAO- and FD-PAW-ED can capture charge oscillation similarly for all energies until the projectile reaches the plane of the graphene. However, for 5~keV and higher, the charge for LCAO-PAW-ED slightly deviates from FD-PAW-ED, resulting in a different final charge on the projectile.~\cite{Kononov2022} Thus we see again how the reliability of our LCAO-PAW-ED simulations starts to suffer for velocities exceeding 4~\AA/fs.

We conducted the same simulations also for C, C${^+}$ and C${^{+2}}$ projectiles to study the effect of more than one valence electron, and compared again against FD-PAW-ED. For C and C${^+}$, the energy loss shows qualitative agreement until the velocity of the projectile reaches 8~\AA/fs, with a nearly constant offset caused by the choice of basis set, whereas for C${^{+2}}$, the two methods start to diverge above 4~\AA/fs. However, total energy conservation again starts to suffer above 4~\AA/fs for all charge states. More details of the C irradiation simulations can be found in the Supplementary information.


\subsection{Computational cost}
\begin{figure}[t!]
    \centering
    \includegraphics[width=0.5\textwidth]{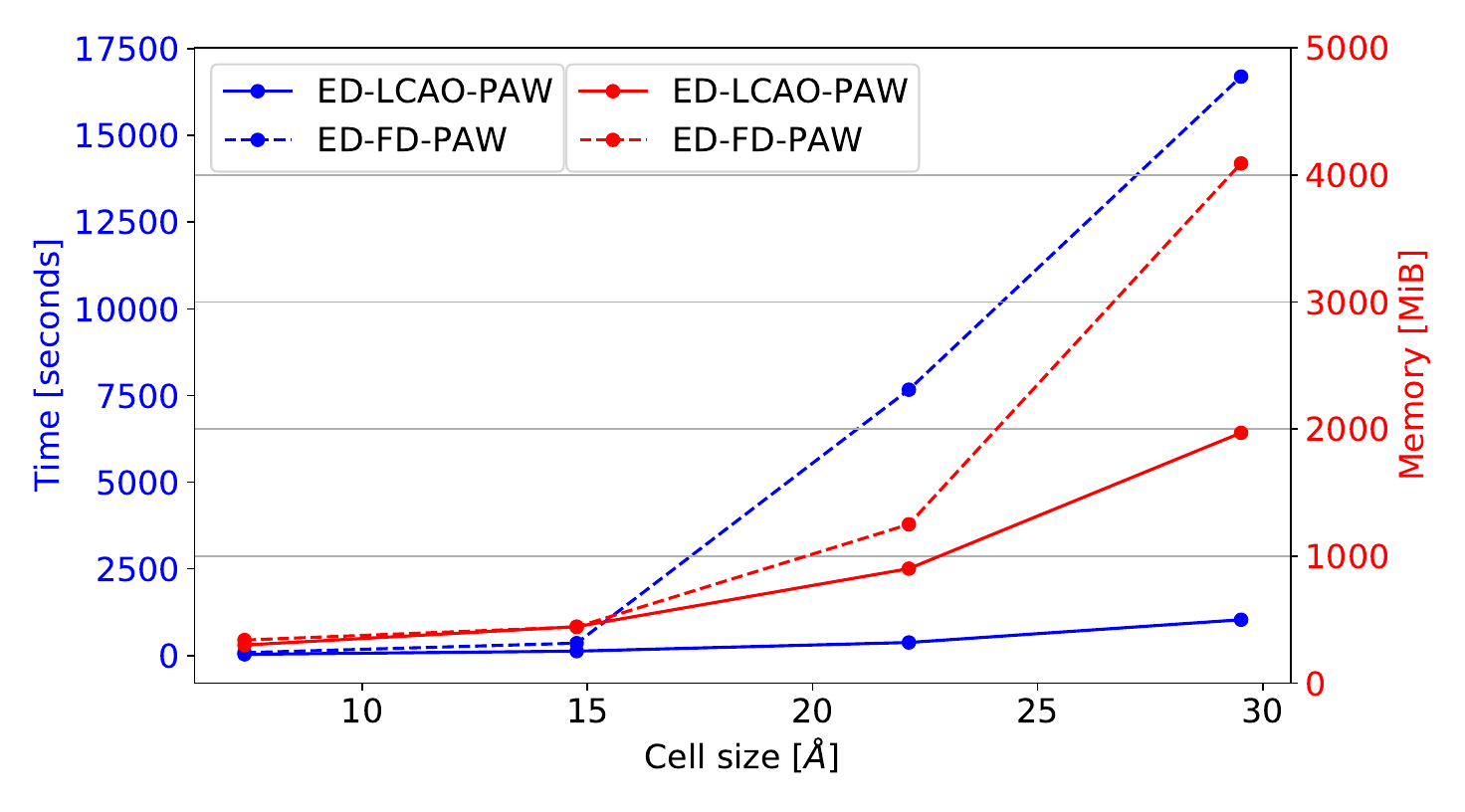}
    \caption{Comparison of computational resources for different methods with a simulation time step of \( \Delta t = 5 \, \text{as} \) for 10 steps parallelized over 96 cores (memory use is plotted per core).}
    \label{fig:MEM_time}
\end{figure}

Finally, we emphasize that despite the limitations of the LCAO-PAW-ED method, it demonstrates substantially greater computational efficiency compared to the FD-PAW-ED method, as shown in Figure~\ref{fig:MEM_time}. For instance, in a simulation supercell of 14.76$\times$14.76$\times$14.76~\AA, 10 timesteps of the LCAO calculation complete in 124.8~s, while FD requires 355.31~s -- nearly a threefold decrease in runtime, both running on 96 cores. For a larger cell, such as 29.52$\times$29.52$\times$29.52~\AA, LCAO performs even more favorably, requiring 1031~s, whereas FD takes 16698~s, corresponding to a speedup of over 16 times. In terms of memory usage, the LCAO approach also offers a significant reduction, using only 1970~MB per core for the largest cell size compared to the FD method's 4090~MB, corresponding to a memory footprint of less than 50\% for larger simulations. Significantly larger cells with more atoms and more vacuum will now be feasible to tackle with LCAO-PAW-ED than was previously possible. These results demonstrate the substantial benefits of the LCAO approach in terms of both runtime and memory, and we believe further optimizations remain possible.

\section{Conclusion}

In summary, our findings highlight the efficacy of Ehrenfest molecular dynamics within the projector augmented-wave formalism of density functional theory using localized basis sets in simulating both molecular systems and particle irradiation. Notably, although the approximate Tomfohr-Sankey integrator is limited to modest atomic velocities, precluding its use for simulating kinetic energies relevant for helium ion microscopy, it is robust against initial charge oscillations of the projectile induced by instantaneous acceleration. 

The notable advantage of LCAO-PAW lies in its substantially lower computational demands compared to the FD-PAW Ehrenfest method, especially in terms of memory use and system-size scaling. Despite its inherent limitations, this method will facilitate simulations of larger systems and longer simulation times, providing a more realistic representation of common experimental conditions. Notably, this may make previously unfeasible simulations with higher charge states potentially tractable.

\section{Supplementary material}

The supplementary material presents additional simulation results for carbon projectiles with charge states of 0, +1, and +2. The effects of projectile charge on interaction dynamics, energy loss, and charge localization are analyzed to distinguish the influence of neutral and ionized projectiles.

\begin{acknowledgments}
We are indebted to the prior work of Ari Ojanperä, and grateful for the advice of GPAW developer Jens Jørgen Mortensen. Funding by the Austrian Science Fund (FWF) [P 36264-N] is acknowledged (V.Z. and T.S.), as well as generous computational resources from the Vienna Scientific Cluster (VSC). T.R. further acknowledges support from the Academy of Finland under Grant No. 332429. A.H.L. acknowledges funding from the European Research Council (ERC) under the European Union's Horizon 2020 research and innovation program Grant No. 951786 (NOMAD CoE).
\end{acknowledgments}

\section*{Data Availability Statement}

The data that support the findings of this study are openly available in the University of Vienna repository Phaidra at http://doi.org/10.25365/phaidra.657, reference number o:2122992.

\appendix

\nocite{*}
\section*{References}
\bibliography{aipsamp}
\end{document}